\documentclass{article}
\usepackage{frascatiphys_net}
\usepackage{hhline}
\usepackage{graphicx}
\begin{document}
\title{
Strangeness measurements of kaon pairs,\\
$CP$ violation and Bell inequalities }
\author{
Reinhold A. Bertlmann and Beatrix C. Hiesmayr \\
{\em Institute for Theoretical Physics, University of Vienna,} \\
{\em Boltzmanngasse 5, A-1090 Vienna, Austria} \\
}
\maketitle
\baselineskip=11.6pt
\begin{abstract}

\begin{center}
\includegraphics[width=115pt,keepaspectratio=true]{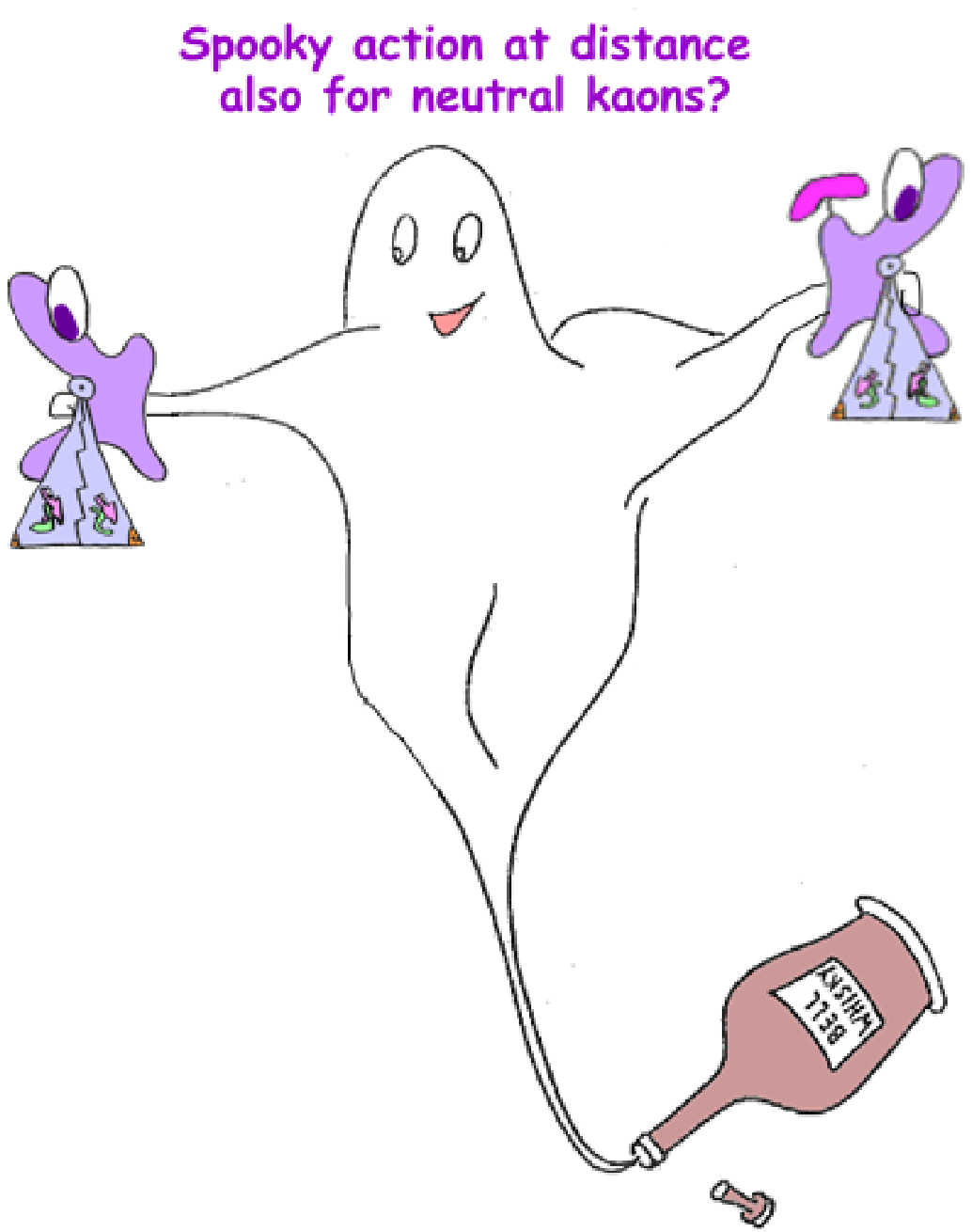}
\end{center}

The nonlocal property of quantum mechanics can be nicely tested in high energy physics;
in particular, the neutral kaon pairs as produced at DA$\Phi$NE, Frascati, are very well
suited. The analogies of kaons as compared to polarized photons or spin--$\frac{1}{2}$
particles ---the kaonic qubit feature--- are reviewed. However, there are also
fundamental differences which occur due to the kaon time evolution and due to internal
symmetries; in particular, the violation of $CP$ symmetry is related to the violation of
Bell inequalities. Two type of Bell inequalities for kaons are presented, one for the
variation of the ``quasi--spin'' and the other for different detection times of the kaon.
\end{abstract}

\baselineskip=14pt

\section{Introduction}

The nonlocality feature of quantum mechanics (QM), as discovered by John Bell in his work
``On the Einstein--Podolsky--Rosen Paradox'' (EPR)\cite{bell}, does not conflict with
Einstein's relativity, thus it cannot be used for superluminal communication.
Nevertheless, Bell's celebrated work\cite{bell,bell2} initiated new physics, like quantum
cryptography\cite{Ekert,DeutschEkert,Hughes,GisinGroup} and quantum
teleportation\cite{Bennett-teleport,ZeilingerTele}, and it triggered a new technology:
quantum information and quantum communication\cite{ZeilingerInfo1,ZeilingerInfo2}. More
about ``from Bell to quantum information'' can be found in the
book\cite{BertlmannZeilinger}.

Of course, it is of great interest to investigate the EPR--Bell correlations of
measurements also for massive systems in particle physics (for a review see, e.g.,
Ref.\cite{BertlmannSchladming}). One of the most exciting systems is the ``strange'' $K^0
\bar K^0$ system in a $J^{PC}=1^{--}$ state\cite{lee,Inglis,Day,Lipkin}, where the
quantum number \textit{strange\-ness} $S=+,-$ plays the role of spin $\Uparrow$ and
$\Downarrow$ of spin--$\frac{1}{2}$ particles or of polarization $V$ and $H$ of photons.
In fact, in comparison to quantum information the kaon can be considered as a ``kaonic
qubit''\cite{BertlmannHiesmayr-KaonicQubit} but due to its specific internal particle
properties (particle--antiparticle oscillation and decay characteristics, symmetry
violation) additional fundamental quantum features ---not occurring in photon systems---
are seen.

Several authors\cite{eberhard,eberhard2,bigi,domenico,Bramon,AncoBramon,BramonGarbarino,Genovese,
BernabeuMavroPapa} suggested already to investigate the $K^0 \bar K^0$  pairs which are produced
at the $\Phi$ resonance, for instance in the $e^+ e^-$--machine DA$\Phi$NE at Frascati. There is
the great chance to test many different aspects of QM, for instance, Bell inequalities and
decoherence models (see, e.g., Ref.\cite{BertlmannSchladming}), the quantum eraser
phenomenon\cite{SBGH1,SBGH6,BramonGarbHiesmayr-Frasc2006} and symmetry
violation\cite{BernabeuMavromatosEtal-Frasc2006}. In particular, local realistic theories (LRT)
have been constructed, which describe the $K^0 \bar K^0$ pairs, as tests versus quantum
mechanics\cite{selleri,SelleriBook,six2,BramonGarbarino2}. However, a general test of LRT versus
QM is usually performed via Bell inequalities, where ---as we shall see--- we have more options.
We may choose either different ``quasi--spins'' of the kaon or different kaon detection times (or
both); they play the role of the different angles in the photon or spin--$\frac{1}{2}$ case. Due
to the kaon decay we have in addition to the \textit{active} measurement procedure the
\textit{passive} measurement. Furthermore, an interesting feature of kaons is $CP$ violation in
the mixing of particle--antiparticle and indeed it is related to the violation of Bell
inequalities.

Besides the kaon system which is an ideal tool to test the amazing features of QM, there
is the $B^0 \bar B^0$ system which is produced to an enormous amount at the asymmetric
$B$--factories at KEK-B\cite{Belle} and at PEP-II\cite{Babar}. A Bell inequality (BI) for
this system\cite{Go} faces, however, with difficulties so that it cannot be considered as
a Bell test refuting local realism. The two main drawbacks are: Firstly, ``active''
measurements ---a necessary requirement for the validity of a BI--- are missing,
therefore one can construct a local realistic model; and secondly, the unitary time
evolution of the unstable quantum state ---the decay property of the meson, which is part
of its nature--- has been ignored (for more detailed criticism, see
Refs.\cite{BBGH,BramonEscribanoGarbarino2004,BramonEscribanoGarbarino2005}).
Nevertheless, the $B^0 \bar B^0$ events, the asymmetry of like-- and unlike--flavor
events for several different times, at KEK-B\cite{Leder} are ideal to test the validity
of the quantum mechanical wavefunction or to confirm the corresponding time dependence of
possible decoherence effects, see
Refs.\cite{BertlmannSchladming,BG1,Dass,BG2,datta,selleribmeson,BG3,
BertlmannDurstbergerHiesmayr2002,CabanPolen2006,KLOEzeta2005,KLOEzeta2006}.

Finally, we want to mention quite different attempts to test QM versus LRT, these are the
positron annihilation
experiments\cite{KasdayUllmannWu1970,KasdayUllmannWu1975,FaraciGNP,WilsonLoweButt,
BrunoAgostinoMaroni,BertoliniDianaScotti}, the proton--proton scattering
experiments\cite{Lamehi-RachtiMittig} and the $\Lambda \bar
\Lambda$\cite{Tornqvist1,Tornqvist2} and $\tau^+ \tau^-$ pair
productions\cite{Privitera,AbelDittmarDreiner}. Unfortunately, all these reactions suffer
by loopholes and are not conclusive as Bell tests (for a detailed discussion, see
Ref.\cite{SelleriBook}).

\section{Kaons as qubits}\label{K-qubits}

Kaons are fantastic quantum systems, we could even say they are selected by Nature to
demonstrate fundamental quantum principles such as:
\begin{itemize}
    \item [$\bullet$] superposition principle
    \item [$\bullet$] oscillation and decay property
    \item [$\bullet$] quasi-spin property.
\end{itemize}
Let us focus on the quantum features which we need in our discussion.

\subsection{{\bf Quantum states of kaons}}\label{K-quantumstates}

Quantum--mechanically we can describe the kaons in the following way. Kaons are
characterized by their \textit{strangeness} quantum number $+1,-1$
\begin{eqnarray}
S|K^0\rangle = + |K^0\rangle \;, \qquad S|\bar K^0\rangle = - |\bar K^0\rangle \;,
\end{eqnarray}
and the combined operation $CP$ gives
\begin{eqnarray}
CP|K^0\rangle = - |\bar K^0\rangle \;, \qquad CP|\bar K^0\rangle = - |K^0\rangle \;.
\end{eqnarray}
It is straightforward to construct the $CP$ eigenstates
\begin{eqnarray}\label{K1K2}
|K_1^0\rangle = \frac{1}{\sqrt{2}}\big\lbrace |K^0\rangle- |\bar K^0\rangle \big\rbrace
\;, \qquad |K_2^0\rangle = \frac{1}{\sqrt{2}}\big\lbrace |K^0\rangle+ |\bar K^0\rangle
\big\rbrace\;,
\end{eqnarray}
a quantum number conserved in strong interactions
\begin{eqnarray}
CP|K_1^0\rangle = + |K_1^0\rangle \;, \qquad CP|K_2^0\rangle = - |K_2^0\rangle \;.
\end{eqnarray}

However, due to weak interactions $CP$ symmetry is \textit{violated} and the kaons decay
in physical states, the short-- and long--lived states, $|K_S\rangle , |K_L\rangle$,
which differ slightly in mass, $\Delta m = m_L - m_S = 3.49 \times 10^{-6}$ eV, but
immensely in their lifetimes and decay modes
\begin{eqnarray}\label{kaonSL}
|K_S\rangle = \frac{1}{N}\big\lbrace p |K^0\rangle-q |\bar K^0\rangle \big\rbrace \;,
\qquad |K_L\rangle = \frac{1}{N}\big\lbrace p |K^0\rangle+q |\bar K^0\rangle \big\rbrace
\;.
\end{eqnarray}
The weights $p=1+\varepsilon$, $\,q=1-\varepsilon,\,$ with $N^2=|p|^2+|q|^2$ contain the
complex $CP$ \textit{violating parameter} $\varepsilon$ with
$|\varepsilon|\approx10^{-3}$. $CPT$ \textit{invariance} is assumed. The short--lived
K--meson decays dominantly into $K_S\longrightarrow 2 \pi$ with a width or lifetime
$\Gamma^{-1}_S\sim\tau_S = 0.89 \times 10^{-10}$ s and the long--lived K--meson decays
dominantly into $K_L\longrightarrow 3 \pi$ with $\Gamma^{-1}_L\sim\tau_L = 5.17 \times
10^{-8}$ s. However, due to $CP$ violation we observe a small amount $K_L\longrightarrow
2 \pi\,$.

In this description the superpositions (\ref{K1K2}) and (\ref{kaonSL}) ---or quite
generally any vector in the 2--dimensional complex Hilbert space of kaons--- represent
kaonic qubit states in analogy to the qubit states in quantum information.

\subsection{{\bf Strangeness oscillation}}\label{strangenessoscillation}

$K_S, K_L$ are eigenstates of a non--Hermitian ``effective mass'' Hamiltonian
\begin{equation}\label{hamiltonian}
H \, = \, M - \frac{i}{2} \,\Gamma
\end{equation}
satisfying
\begin{equation}
H \,|K_{S,L}\rangle \; = \; \lambda_{S,L} \,|K_{S,L}\rangle \qquad \textrm{with} \qquad
\lambda_{S,L} \, = \, m_{S,L} - \frac{i}{2} \,\Gamma_{S,L} \;.
\end{equation}
Both mesons $K^0$ and $\bar K^0$ have transitions to common states (due to $CP$
violation) therefore they mix, that means they \textit{oscillate} between $K^0$ and $\bar
K^0$ before decaying. Since the decaying states evolve ---according to the
Wigner--Weisskopf approximation--- exponentially in time
\begin{equation}\label{Wigner--Weisskopf}
| K_{S,L} (t)\rangle \; = \; e^{-i \lambda_{S,L} t} | K_{S,L} \rangle \;,
\end{equation}
the subsequent time evolution for $K^0$ and $\bar K^0$ is given by
\begin{eqnarray}\label{K-time-evolution}
| K^0(t) \rangle = g_{+}(t) | K^0 \rangle  + \frac{q}{p} g_{-}(t) | \bar K^0 \rangle \;,
\quad | \bar K^0(t) \rangle = \frac{p}{q} g_{-}(t) | K^0 \rangle + g_{+}(t) | \bar K^0
\rangle
\end{eqnarray}
with
\begin{equation}\label{g+-}
g_{\pm}(t) \, = \, \frac{1}{2} \left[ \pm e^{-i \lambda_S t} + e^{-i \lambda_L t} \right]
\;.
\end{equation}
Supposing that a $K^0$ beam is produced at $t=0$, e.g. by strong interactions, then the
probability for finding a $K^0$ or $\bar K^0$ in the beam is calculated to be
\begin{eqnarray}
\left| \langle K^0 | K^0(t) \rangle \right|^2 &=& \frac{1}{4} \big\lbrace e^{-\Gamma_S t}
+ e^{-\Gamma_L t} + 2 \, e^{-\Gamma t}
\cos(\Delta m t)\big\rbrace \;, \nonumber\\
\left| \langle \bar K^0 | K^0(t) \rangle \right|^2 &=& \frac{1}{4} \frac{|q|^2}{|p|^2}
\big\lbrace e^{-\Gamma_S t} + e^{-\Gamma_L t} - 2 \, e^{-\Gamma t} \cos(\Delta m
t)\big\rbrace \, ,
\end{eqnarray}
with $\Delta m=m_L-m_S\,$ and $\,\Gamma = \frac{1}{2}(\Gamma_L+\Gamma_S)\,$.

The $K^0$ beam oscillates with frequency $\Delta m / 2\pi$, where $\Delta m \, \tau_S =
0.47$. The oscillation is clearly visible at times of the order of a few $\tau_S$, before
all $K_S$'s have died out leaving only the $K_L$'s in the beam. So in a beam which
contains only $K^0$ mesons at the beginning $t=0$ there will occur $\bar K^0$ far from
the production source through its presence in the $K_L$ meson.

\subsection{{\bf Quasi--spin of kaons and analogy to photons}}\label{quasispin}

In comparison with spin--$\frac{1}{2}$ particles, or with photons having the polarization
directions V (vertical) and H (horizontal), it is very instructive to characterize the
kaons by a \textit{quasi--spin} (for details see Ref.\cite{BertlmannHiesmayr2001}). We
can regard the two states $| K^0 \rangle$ and $| \bar K^0 \rangle$ as the quasi--spin
states up $\mid\uparrow\rangle$ and down $\mid\downarrow\rangle$ and can express the
operators acting in this quasi--spin space by Pauli matrices. So we identify the
strangeness operator $S$ with the Pauli matrix $\sigma_3$, the $CP$ operator with
($-\sigma_1$) and for describing $CP$ violation we also  need $\sigma_2$. In fact, the
Hamiltonian (\ref{hamiltonian}) then has the form
\begin{equation}
H \, = \, a\cdot \mathbf{1} + \vec b \cdot \vec \sigma \;,
\end{equation}
with
\begin{eqnarray}
b_1 = b \cos \alpha, \quad b_2 = b \sin \alpha, \quad b_3 = 0 \;, \nonumber\\
a = \frac{1}{2}(\lambda_L + \lambda_S), \quad b = \frac{1}{2}(\lambda_L - \lambda_S) \;,
\end{eqnarray}
and the angle $\alpha$ is related to the $CP$ violating parameter $\varepsilon$ by
\begin{equation}
e^{i\alpha} \, = \, \frac{1-\varepsilon}{1+\varepsilon} \;.\\
\end{equation}

\vspace{0.3cm}

Summarizing, we have the following kaonic--photonic analogy:

\vspace{0.2cm}

\begin{center}
\renewcommand{\arraystretch}{1.7}
\begin{tabular}{||c|c|c||}
\hhline{|t:===:t|}
  $\;$\textbf{neutral kaon}$\;$& $\;$ \textbf{quasi--spin}$\;$ &\textbf{photon}\\
\hhline{|:===:|}
  $| K^0\rangle$ & $| \uparrow\rangle_z$ & $| V\rangle$ \\
  $ |\bar K^0\rangle$ &  $| \downarrow\rangle_z$ & $| H\rangle$ \\
\hhline{|:===:|}
  $| K_1^0\rangle$ &  $| \nwarrow\rangle$ &  $| -45^0\rangle =
  \frac{1}{\sqrt{2}}(| V\rangle - | H\rangle)$ \\
  $| K_2^0\rangle$ & $| \nearrow\rangle$ &  $| +45^0\rangle =
  \frac{1}{\sqrt{2}}(| V\rangle + | H\rangle)$ \\
\hhline{|:===:|}
  $| K_S\rangle$ & $| \rightarrow\rangle_y$ & $| L\rangle =
  \frac{1}{\sqrt{2}}(| V\rangle - i| H\rangle)$ \\
  $| K_L\rangle$ & $| \leftarrow\rangle_y$ & $| R\rangle =
  \frac{1}{\sqrt{2}}(| V\rangle + i| H\rangle)$ \\
\hhline{|b:===:b|}
\end{tabular}
\renewcommand{\arraystretch}{1}
\end{center}

A good \textit{optical analogy} to the phenomenon of strangeness oscillation can  be
achieved by using the physical effect of birefringence in optical fibers which leads to
the rotation of polarization directions. Thus $H$ (horizontal) polarized light is rotated
after some distance into $V$ (vertical) polarized light, and so on. On the other hand,
the decay of kaons can be simulated by polarization dependent losses in optical fibres,
where one state has lower losses than its orthogonal state\cite{GisinGo}.
\\
\\
The description of kaons as qubits reveals close analogies to photons but also deep
physical differences. Kaons oscillate, they are massive, they decay and can be
characterized by symmetries like $CP$. Even though some kaon features, like oscillation
and decay, can be mimicked by photon experiments (see Ref.\cite{GisinGo}), they are
inherently different since they are intrinsic properties of the kaon given by Nature.

\subsection{{\bf Measurement procedures}}\label{measurementprocedures}

For neutral kaons there exist two physical alternative bases, accordingly we have two
observables for the kaons, namely the projectors to the two bases. The first basis is the
strangeness eigenstate basis $\{| K^0\rangle, |\bar K^0 \rangle\}$, it can be measured by
inserting along the kaon trajectory a piece of ordinary matter, which corresponds to an
\textit{active} measurement of strangeness. Due to strangeness conservation of the strong
interactions the incoming state is projected either onto $K^0$ by $K^0 p\rightarrow K^+
n$ or onto $\bar K^0$ by $\bar K^0 p\rightarrow \Lambda \pi^+$, $\bar K^0 n\rightarrow
\Lambda \pi^0$ or $\bar K^0 n\rightarrow K^- p$. Here nucleonic matter plays the same
role as a two channel analyzer for polarized photon beams.

Alternatively, the strangeness content of neutral kaons can be determined by observing
their semileptonic decay modes, eq.(\ref{semileptonic-decays}).

Obviously, the experimenter has no control of the kaon decay, neither of the mode nor of
the time. The experimenter can only sort at the end of the day all observed events in
proper decay modes and time intervals. We call this procedure opposite to the
\textit{active} measurement described above a \textit{passive} measurement procedure of
strangeness.

The second basis $\{K_S,K_L\}$ consists of the short-- and long--lived states having well
defined masses $m_{S(L)}$ and decay widths $\Gamma_{S(L)}$. We have seen that it is the
appropriate basis to discuss the kaon propagation in free space, because these states
preserve their own identity in time, eq.(\ref{Wigner--Weisskopf}). Due to the huge
difference in the decay widths the $K_S$'s decay much faster than the $K_L$'s. Thus in
order to observe if a propagating kaon is a $K_S$ or $K_L$ at an instant time $t$, one
has to detect at which time it subsequently decays. Kaons which are observed to decay
before $\simeq t + 4.8\, \tau_S$ have to be identified as $K_S$'s, while those surviving
after this time are assumed to be $K_L$'s. Misidentifications reduce only to a few parts
in $10^{-3}$ (see Refs.\cite{SBGH1,SBGH6}). Note that the experimenter doesn't care about
the specific decay mode, she/he records only a decay event at a certain time. We call
this procedure an \textit{active} measurement of lifetime.

Since the neutral kaon system violates the $CP$ symmetry (recall Section
\ref{K-quantumstates}) the mass eigenstates are not strictly orthogonal, $\langle
K_S|K_L\rangle\neq 0$. However, neglecting $CP$ violation ---remember it is of the order
of $10^{-3}$--- the $K_S$'s are identified by a $2\pi$ final state and $K_L$'s by a
$3\pi$ final state. We call this procedure a \textit{passive} measurement of lifetime,
since the kaon decay times and decay channels used in the measurement are entirely
determined by the quantum nature of kaons and cannot be in any way influenced by the
experimenter. It is assumed that \textit{active} and \textit{passive} measurements have
the same amount of misidentifications.\\

The important message for testing Bell inequalities which we are going to discuss in the
next section is:
\begin{itemize}
    \item [$\bullet$] The \textit{active} measurement procedures are a necessary requirement
    for the validity of a BI.
\end{itemize}

\section{Entangled kaons, Bell inequalities, $CP$ violation}\label{BI}

Having discussed kaons as qubit states and their analogy to photons we consider next two
qubit states. A two qubit system of kaons is a general superposition of the 4 states
\mbox{$\{| K^0 \rangle \otimes | K^0 \rangle$}, \mbox{$| K^0 \rangle \otimes | \bar K^0
\rangle ,$} $| \bar K^0 \rangle \otimes | K^0 \rangle , | \bar K^0 \rangle \otimes | \bar
K^0 \rangle\}\,$.

\subsection{{\bf Entanglement}}\label{entanglement}

Interestingly, also for strange mesons entangled states can be obtained, in analogy to
the entangled spin up and down pairs, or H and V polarized photon pairs. Such states are
produced by $e^+ e^-$--colliders through the reaction $e^+ e^- \to \Phi \to K^0 \bar
K^0$, in particular at DA$\Phi$NE in Frascati, or they are produced in $p\bar
p$--collisions, like, e.g., at LEAR at CERN\cite{CPLEAR}. There, a $K^0 \bar K^0$ pair is
created in a $J^{PC}=1^{--}$ quantum state and thus antisymmetric under $C$ and $P$, and
is described at the time $t=0$ by the entangled state
\begin{eqnarray}\label{entangledK0}
| \psi (t=0) \rangle &=&\frac{1}{\sqrt{2}} \left\{ | K^0 \rangle_l \otimes | \bar K^0
\rangle _r - | \bar K^0 \rangle _l \otimes | K^0 \rangle _r \right\}\,,\nonumber\\
&=& \frac{N_{SL}}{\sqrt{2}}\left\{ | K_S \rangle_l \otimes | K_L \rangle _r - | K_L
\rangle _l \otimes | K_S \rangle _r \right\}\,,
\end{eqnarray}
with $N_{SL}=\frac{N^2}{2pq}$, in complete analogy to the entangled photon case
\begin{eqnarray}\label{entangled-photon}
| \psi \rangle &=&\frac{1}{\sqrt{2}} \left\{ | V \rangle_l \otimes | H
\rangle _r - | H \rangle _l \otimes | V \rangle _r \right\}\,,\nonumber\\
&=& \frac{1}{\sqrt{2}}\left\{ | L \rangle_l \otimes | R \rangle _r - | R \rangle _l
\otimes | L \rangle _r \right\}\,.
\end{eqnarray}
The neutral kaons fly apart and are detected on the left ($l$) and right ($r$) hand side
of the source. Of course, during their propagation the $K^0 \bar K^0$ pairs oscillate and
the $K_S, K_L$ states decay. This is an important difference to the case of photons which
are stable.

Let us measure \textit{actively} at time $t_l$ a $K^0$ meson on the left hand side and at
time $t_r$ a $K^0$ or a $\bar K^0$ on the right hand side then we find an EPR--Bell
correlation analogously to the entangled photon case with polarization V--V or V--H.
Assuming for simplicity stable kaons ($\Gamma _S = \Gamma _L = 0$) then we get the
following result for the quantum probabilities
\begin{eqnarray}
P(K^0,t_l;K^0,t_r) &=& P(\bar K^0,t_l;\bar K^0,t_r) \; = \; \frac{1}{4}
\big\lbrace 1 - \cos(\Delta m(t_l - t_r))\big\rbrace \;,\nonumber\\
P(K^0,t_l;\bar K^0,t_r) &=& P(\bar K^0,t_l;K^0,t_r) \; = \; \frac{1}{4} \big\lbrace 1 +
\cos(\Delta m(t_l - t_r))\big\rbrace \;,
\end{eqnarray}
which is the analogy to the probabilities of finding simultaneously two entangled photons
along two chosen directions $\vec \alpha$ and $\vec \beta$
\begin{eqnarray}
P(\vec \alpha,V ;\vec \beta,V) &=& P(\vec \alpha,H ;\vec \beta,H) \; = \; \frac{1}{4}
\big\lbrace 1 - \cos 2(\alpha - \beta) \big\rbrace \;,\nonumber\\
P(\vec \alpha,V ;\vec \beta,H) &=& P(\vec \alpha,H;\vec \beta,V) \; = \; \frac{1}{4}
\big\lbrace 1 + \cos 2(\alpha - \beta) \big\rbrace \;.
\end{eqnarray}
Thus we observe a \textit{perfect analogy} between times $\Delta m(t_l - t_r)$ and angles
$2(\alpha - \beta)$.\\

Alternatively, we also can fix the time and vary the quasi--spin of the kaon, which
corresponds to a rotation in quasi--spin space analogously to the rotation of
polarization of the photon
\begin{eqnarray}
\mid k\rangle = a\mid K^0\rangle + b\mid\bar K^0\rangle \quad \longleftrightarrow \quad
\mid \alpha, \phi; V\rangle = \cos\alpha \mid V \rangle +
    \sin\alpha \,e^{i \phi}\mid H \rangle \;.
\end{eqnarray}

Note that the weights $a, b$ are not independent and not all kaonic superpositions are
realized in Nature in contrast to photons.

\vspace{0.3cm}

Depicting the kaonic--photonic analogy we have:

\vspace{0.3cm}

\begin{center}
\small{\hspace{0.2cm}\textbf{kaon propagation} \hspace{1.38cm} \textbf{photon propagation}\\
\setlength{\unitlength}{1cm}

\vspace{0.45cm}

\begin{picture}(10.5,1)(-1.75,-0.6)
    \put(0.48,0){\vector(-1,0){1.5}}
    \put(0.8,0){\vector(1,0){1.5}}
    \put(0.65,0){\circle{0.3}}
    \put(-1.5,0.15){$K^0/K_S$}
    \put(1.7,0.15){$\bar K^0/K_L$}
    \put(3.9,0.15){$V/L$}
    \put(7.1,0.15){$H/R$}
    \put(-0.1,-0.6){\footnotesize{Bell state}}
    \put(-1.2,-0.5){{\footnotesize{left}}}
    \put(1.9,-0.5){{\footnotesize{right}}}
    \put(3.9,-0.5){{\footnotesize{Alice}}}
    \put(7.2,-0.5){{\footnotesize{Bob}}}
    \put(5.68,0){\vector(-1,0){1.5}}
    \put(6.0,0){\vector(1,0){1.5}}
    \put(5.85,0){\circle{0.3}}
    \put(5.4,-0.6){\footnotesize{Bell state}}
\end{picture}

\vspace{0.25cm}

{\footnotesize{\hspace{-3.2cm}$\bullet$ $K^0\bar K^0$ oscillation  \hspace{2.2cm}
$\bullet$ stable
\\ \hspace{-7.35cm} $\bullet$ $K_S$, $K_L$ decay
}}}
\end{center}

\vspace{0.3cm}

\subsection{{\bf Bell inequality for quasi--spin variation}}\label{Bell-inequality-quasispin}

Consequently, for establishing a BI for kaons we have the option:
\begin{enumerate}
\item[$\bullet$] varying the quasi--spin --- fixing time
\item[$\bullet$] fixing the quasi--spin --- varying time.
\end{enumerate}

Let us begin with a BI for certain quasi--spins (first option) and demonstrate that its
violation is related to a symmetry violation in particle physics. In
Ref.\cite{Nagata,Unnikrishnan} it was shown that symmetries quite generally may constrain
local realistic theories.

For a BI we need $3$ different ``quasi--spins'' -- the ``Bell angles'' -- and we may
choose the $H$, $S$ and $CP$ eigenstates: $|K_S\rangle\,,|\bar K^0\rangle\,$ and
$|K_1^0\rangle \,$.

Denoting the probability of measuring the short--lived state $K_S$ on the left hand side
and the anti--kaon $\bar K^0$ on the right hand side, both at the time $t=0$, by
$P(K_S,\bar K^0)$, and analogously the probabilities $P(K_S,K_1^0)$ and $P(K_1^0,\bar
K^0)$ we can easily derive under the usual hypothesis of Bell's locality the following
\textit{Wigner--like Bell inequality}\cite{Uchiyama,BGH-CP}
\begin{equation}\label{UchiyamaBI}
P(K_S,\bar K^0)\; \leq\; P(K_S,K_1^0) + P(K_1^0,\bar K^0) \;.
\end{equation}
BI (\ref{UchiyamaBI}) is rather formal because it involves the unphysical $CP$--even
state $| K^0_1 \rangle$, but -- and this is now important -- it implies an inequality on
a \textit{physical} quantity, the $CP$ violation parameter. Inserting the quantum
amplitudes
\begin{equation}
\langle \bar{K}^0\mid K_S\rangle=-\frac{q}{N}\,,\;\quad \langle \bar{K}^0\mid
K_1^0\rangle = - \frac{1}{\sqrt{2}}\,,\;\quad \langle K_S\mid K_1^0\rangle =
\frac{1}{\sqrt{2}N}(p^*+q^*)\,,
\end{equation}
and optimizing the inequality we can convert (\ref{UchiyamaBI}) into an inequality for
the complex kaon transition coefficients $p,q$
\begin{eqnarray}\label{inequalpq}
|\,p\,| &\leq& |\,q\,| \;.
\end{eqnarray}
It's amazing, inequality (\ref{inequalpq}) is \textit{experimentally testable}! How does
it work?

\subsection{{\bf Semileptonic decays}}\label{sect-semileptonicdecays}

Let us consider the semileptonic decays of the kaons. The strange quark $s$ decays weakly
as constituent of $\bar K^0\,$ (see Fig.\ref{s-quark-decay-fig}):

\begin{figure}[t]
 \vspace{2.5cm}
\includegraphics{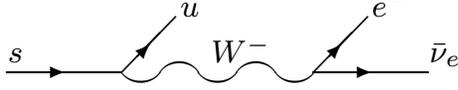}
 \caption{\it
      Strange quark decays weakly.
    \label{s-quark-decay-fig} }
\end{figure}


Due to the quark content $K^0(\bar s d)$ and $\bar K^0(s \bar d)$ have the following
decays:
\begin{eqnarray}\label{semileptonic-decays}
K^0(d\bar{s}) \;&\longrightarrow&\; \pi^-(d\bar{u})\;\; l^+\;\nu_l \qquad
\textrm{where} \qquad \bar{s} \;\longrightarrow\; \bar{u}\;\; l^+\;\nu_l \nonumber \\
\bar{K}^0(\bar{d}s) \;&\longrightarrow&\; \pi^+(\bar{d}u)\;\; l^-\;\bar{\nu}_l \qquad
\textrm{where} \qquad  s \;\longrightarrow\; u\;\; l^-\;\bar{\nu}_l \;,
\end{eqnarray}
with $l=\mu, e\,$. When studying the leptonic charge asymmetry
\begin{eqnarray}\label{asymlept}
\delta &=& \frac{\Gamma(K_L\rightarrow \pi^- l^+ \nu_l) - \Gamma(K_L\rightarrow \pi^+ l^-
\bar \nu_l)}{\Gamma(K_L\rightarrow \pi^- l^+ \nu_l) + \Gamma(K_L\rightarrow \pi^+ l^-
\bar \nu_l)} \;,
\end{eqnarray}
we notice that $l^+$ and $l^-$ tag $K^0$ and $\bar K^0$, respectively, in the $K_L$
state, and the leptonic asymmetry (\ref{asymlept}) is expressed by the probabilities
$|p|^2$ and $|q|^2$ of finding a $K^0$ and a $\bar K^0$, respectively, in the $K_L$ state
\begin{eqnarray}
\delta &=& \frac{|p|^2-|q|^2}{|p|^2+|q|^2} \;.
\end{eqnarray}

Returning to inequality (\ref{inequalpq}) we find consequently the bound
\begin{eqnarray}\label{inequaldelta}
\delta &\leq& 0
\end{eqnarray}
for the leptonic charge asymmetry which measures $CP$ violation.

Experimentally, however, the asymmetry is nonvanishing\cite{ParticleData}
\begin{equation}\label{deltaexp}
\delta = (3.27 \pm 0.12)\cdot 10^{-3} \;.
\end{equation}
What we find is that bound (\ref{inequaldelta}), dictated by BI (\ref{UchiyamaBI}), is in
contradiction to the experimental value (\ref{deltaexp}) which is definitely positive.

On the other hand, we can replace $\bar K^0$ by $K^0$ in the BI (\ref{UchiyamaBI}) and
obtain the reversed inequality $\delta \geq 0$ so that respecting all possible BI's leads
to strict equality $\delta = 0$, $CP$ conservation, in contradiction to experiment.\\

\textit{Conclusion}: The premises of LRT are \textit{only} compatible with strict $CP$
conservation in $K^0 \bar K^0$ mixing. Conversely, $CP$ violation in $K^0 \bar K^0$
mixing, no matter which sign the experimental asymmetry (\ref{asymlept}) actually has,
always leads to a \textit{violation} of a BI and in consequence rules out a local
realistic theory for the description of a $K^0 \bar K^0$ system!\\

\textit{Remark}: We believe that this connection between symmetry violation and BI
violation is not just accidental for the $CP$ symmetry case but is more fundamental and
should be observed in case of other symmetries as well.\\

\subsection{{\bf Bell inequality for time variation}}\label{Bell-inequality-time}

Bell inequalities by fixing the quasi--spin and varying the time we have studied already
in detail in Refs.\cite{BertlmannHiesmayr2001,BBGH,ThesisTrixi,Hiesmayr2006}. As we
emphasized in a \textit{unitary} time evolution also the decay states are involved, in
fact, in the following way.

The complete time evolution of the kaon states is given by a \textit{unitary} operator
$U(t,0)$ whose effect can be written as\cite{BellSteinberger,ghirardi91}
\begin{eqnarray}\label{timeevolution}
U(t,0)\; |K_{S,L}\rangle &=& e^{-i \lambda_{S,L} t}\;|K_{S,L}\rangle +
|\Omega_{S,L}(t)\rangle \,,
\end{eqnarray}
where $|\Omega_{S,L}(t)\rangle$ denotes the state of all decay products. The norm
decrease of the state $| K_{S,L}(t) \rangle$ must be compensated by the increase of the
norm of the final states, i.e., $\langle \Omega_{S,L}(t)|\Omega_{S,L}(t)\rangle =
1-e^{-\Gamma_{S,L}\, t}$ and $\langle \Omega_L(t)|\Omega_S(t)\rangle = \langle
K_L|K_S\rangle (1-e^{i \Delta m \,t}e^{-\Gamma t}), \,\langle K_{S,L}|\Omega_S(t)\rangle
= \langle K_{S,L}|\Omega_L(t)\rangle=0 \,$.

Let us start at time $t=0$ with an entangled state of kaon pairs given in the $K_S K_L$
basis choice (recall eq.\ref{entangledK0})
\begin{eqnarray}\label{entangled-KsKl}
| \psi (t=0) \rangle &=& \frac{N_{SL}}{\sqrt{2}}\left\{ | K_S \rangle_l \otimes | K_L
\rangle _r - | K_L \rangle _l \otimes | K_S \rangle _r \right\}\,.
\end{eqnarray}
Then we get the state at time $t$ from (\ref{entangled-KsKl}) by applying the unitary
operator
\begin{eqnarray}\label{U(t)unitary}
U(t,0) &=& U_l(t,0) \cdot U_r(t,0) \, ,
\end{eqnarray}
where the operators $U_l(t,0)$ and $U_r(t,0)$ act on the space of the left and of the
right mesons according to the time evolution (\ref{timeevolution}).

For the quantum mechanical probabilities for detecting, or not detecting, a specific
quasi--spin state on the left side, say $|\bar K^0\rangle_l$, and on the right side
$|\bar K^0\rangle_r$ of the source we need the projection operators
\begin{eqnarray}
P_{l,r}(\bar K^0) \, = \, |\bar K^0\rangle \langle \bar K^0|_{l,r} \qquad \mbox{and}
\qquad Q_{l,r}(\bar K^0) \, = \, \mathbf{1} - P_{l,r}(\bar K^0) \, .
\end{eqnarray}
Starting from the initial state (\ref{entangled-KsKl}) the unitary time evolution
(\ref{U(t)unitary}) provides the state at a time $t_r$
\begin{eqnarray}
|\psi(t_r)\rangle &=& U(t_r,0)|\psi(t=0)\rangle \; = \; U_l(t_r,0) U_r(t_r,0)
|\psi(t=0)\rangle \, .
\end{eqnarray}
Measuring now $\bar K^0$ at $t_r$ on the right side means that we project onto the state
\begin{eqnarray}\label{psi-r}
|\tilde{\psi}(t_r)\rangle &=& P_r(\bar K^0) |\psi(t_r)\rangle \, ,
\end{eqnarray}
and state (\ref{psi-r}) evolves until $t_l$ when we measure next a $\bar K^0$ on the left
side
\begin{eqnarray}\label{evolutionexact}
|\tilde{\psi}(t_l, t_r)\rangle &=& P_l(\bar K^0) U_l(t_l,t_r) P_r(\bar K^0)
|\psi(t_r)\rangle \, .
\end{eqnarray}
The probability of the joint measurement is given by the squared norm of the state
(\ref{evolutionexact}) and coincides with the norm of the state
\begin{eqnarray}\label{evolutionfactorized}
|\psi(t_l,t_r)\rangle &=& P_l(\bar K^0) P_r(\bar K^0) U_l(t_l,0) U_r(t_r,0)
|\psi(t=0)\rangle \, ,
\end{eqnarray}
which corresponds to a factorization of the eigentimes $t_l$ and $t_r$.

We calculate the quantum mechanical probability $P_{\bar K^0,\bar K^0}(Y, t_l; Y, t_r)$
for finding a $\bar K^0$ at $t_l$ on the left side {\it and} a $\bar K^0$ at $t_r$ on the
right side, and the probability $P_{\bar K^0,\bar K^0}(N, t_l; N, t_l)$ for finding $no$
such kaons by the following norms (and similarly the probability $P_{\bar K^0,\bar
K^0}(Y, t_l; N, t_r)$)
\begin{eqnarray}\label{ProbabilitiesYN}
P_{\bar K^0,\bar K^0}(Y, t_l; Y, t_r) &=& ||P_l(\bar K^0) P_r(\bar K^0) U_l(t_l,0)
U_r(t_r,0)|\psi(t=0)\rangle||^2 \\
P_{\bar K^0,\bar K^0}(N, t_l; N, t_r) &=& ||Q_l(\bar K^0) Q_r(\bar K^0) U_l(t_l,0)
U_r(t_r,0)|\psi(t=0)\rangle||^2 \\
P_{\bar K^0,\bar K^0}(Y, t_l; N, t_r) &=& ||P_l(\bar K^0) Q_r(\bar K^0) U_l(t_l,0)
U_r(t_r,0) |\psi(t=0)\rangle||^2 .
\end{eqnarray}
Then the expectation value for measuring the antikaons is expressed by
\begin{eqnarray}\label{meanvalueprob}
E_{\bar K^0,\bar K^0}(t_l,t_r) = -1 + 2\, \big\lbrace P_{\bar K^0,\bar K^0}(Y, t_l; Y,
t_r) + P_{\bar K^0,\bar K^0}(N, t_l; N, t_r)\big\rbrace \,,
\end{eqnarray}
and with expression (\ref{meanvalueprob}) Bell inequalities are constructed.\\

For our purpose we use a BI in the familiar expression of Clauser, Horne, Shimony, Holt
(CHSH)\cite{CHSH} which in terms of time variation can be formulated in the following
way\cite{BertlmannHiesmayr2001,ghirardi91}. Defining the function
\begin{eqnarray}\label{S-function}
S(t_1,t_2,t_3,t_4) &=& | E_{\bar K^0,\bar K^0}(t_1,t_2) - E_{\bar K^0,\bar
K^0}(t_1,t_3)|\nonumber\\
&& + |E_{\bar K^0,\bar K^0}(t_4,t_2) + E_{\bar K^0,\bar K^0}(t_4,t_3)| \,,
\end{eqnarray}
the CHSH--Bell inequality is given by
\begin{eqnarray}\label{CHSH-inequality}
S(t_1,t_2,t_3,t_4) \; \leq \; 2 \,,
\end{eqnarray}
where the value $2$ is the maximum satisfied by any LRT.\\

The question is now whether inequality (\ref{CHSH-inequality}) can be violated in the kaon case.
As we know\cite{BertlmannHiesmayr2001,ThesisTrixi,ghirardi91} the four Bell states
($\psi^{\mp}\sim K_S K_L \pm K_L K_S  \,,\, \phi^{\mp}\sim K_S K_S \mp K_L K_L$) which are maximal
entangled do not violate inequality (\ref{CHSH-inequality}). The reason is that the internal
physical parameters, the ratio oscillation to decay, $\Delta m / \Gamma$, is experimentally about
$1$ whereas for a violation a value of $2$ is necessary for the $\psi^{\mp}$ states and a smaller
value of about $1.7$ for the $\phi^{\mp}$ states.

\vspace{0.1cm}

A recent investigation\cite{Hiesmayr2006} of a quite general initial state
\begin{eqnarray}\label{generalinitialstate}
|\psi(0)\rangle &=& r_1 e^{i \phi_1} |K_S\rangle_l \otimes |K_S\rangle_r \,+\, r_2
e^{i \phi_2} |K_S\rangle_l \otimes |K_L\rangle_r\nonumber\\
&& \,+\, r_3 e^{i \phi_3} |K_L\rangle_l \otimes |K_S\rangle_r \,+\, r_4 e^{i \phi_4}
|K_L\rangle_l \otimes |K_L\rangle_r \;,
\end{eqnarray}
(with $r_1^2+r_2^2+r_3^2+r_4^2=1$) providing the general expectation value
\begin{eqnarray}\label{expectationvalue}
\lefteqn{E_{\bar K^0,\bar K^0}(t_l,t_r)\; =\; 1+ r_1^2\, e^{-\Gamma_S (t_l+t_r)}+r_2^2\,
e^{-\Gamma_S t_l-\Gamma_L t_r}+r_3^2\, e^{-\Gamma_L t_l-\Gamma_S t_r}}\nonumber\\
&&+r_4^2\, e^{-\Gamma_L (t_l+t_r)} - r_1^2\, (e^{-\Gamma_S t_l}+ e^{-\Gamma_S
t_r})-r_2^2\, (e^{-\Gamma_S t_l}+ e^{-\Gamma_L t_r})\nonumber\\
&&- r_3^2\, (e^{-\Gamma_L t_l}+ e^{-\Gamma_S t_r}) - r_4^2\, (e^{-\Gamma_L t_l}+
e^{-\Gamma_L t_r})\nonumber\\
&&+ 2\,  r_1 r_2\, (1-e^{-\Gamma_S t_l}) \cos(\Delta m t_r+\phi_1-\phi_2)\, e^{-\Gamma
t_r}\nonumber\\
&&+2\, r_1 r_3\, \cos(\Delta m t_l+\phi_1-\phi_3)\, e^{-\Gamma t_l}\,(1-e^{-\Gamma_S
t_r})\nonumber\\
&&+ 2\,  r_2 r_4\,  \cos(\Delta m t_l+\phi_2-\phi_4)\, e^{-\Gamma t_l}\, (1-e^{-\Gamma_L
t_r})\nonumber\\
&&+2\, r_3 r_4\, (1-e^{-\Gamma_L t_l})\,\cos(\Delta m t_r+\phi_3-\phi_4)\,
 e^{-\Gamma t_r}\nonumber\\
&&+ 2\, r_1 r_4 \cos(\Delta m (t_l+t_r)+\phi_1-\phi_4)\, e^{-\Gamma (t_l+t_r)}\nonumber\\
&&+2\, r_2 r_3\, \cos(\Delta m (t_l-t_r)+\phi_2-\phi_3)\, e^{-\Gamma (t_l+t_r)} \;,
\end{eqnarray}
shows that for a certain parameter choice the CHSH--Bell inequality
(\ref{CHSH-inequality}) is indeed \textit{violated}!

The $S$--function value turns out to be $S=2.12$ for the parameter choice: all phases
$\phi_i=0$ and $r_1=-0.834, \, r_2=r_3=0.245$ and times $t_1=t_2=0, \,
t_3=t_4=5.6\tau_S$;\, and $S=2.16$ for the choice: $\phi_1=-0.275,\,
\phi_2=\phi_3=-0.678$ and $r_1=-0.782,\, r_2=r_3=-0.146$ and times $t_1=t_2=1.6\tau_S, \,
t_3=t_4=0\,$. (The numerical optimization procedure does not guarantee a global maximum).

\textit{Conclusion}: There exist initial states for kaons that ---by respecting the unitary time
evolution, the decay property--- violate a Bell inequality and are therefore nonlocal, although
not maximal entangled, which agrees with the qutrit results of
Refs.\cite{GisinPRL2005,GisinPRA2002}. It shows that nonlocality and entanglement are \textit{not}
the same features of QM. The question remains, however, how to produce the initial state
(\ref{generalinitialstate}) with the parameter values given above, e.g., at DA$\Phi$NE.

\section{Conclusions}

Kaons are ideal objects to test the fundamental principles of quantum mechanics, in
particular the entanglement or nonlocality properties of kaon pairs, which are of great
interest in connection with the physics of quantum communication and quantum information.
In fact, in analogy to polarized photons the kaons can be considered as qubits as well
but ---due to their internal symmetries and time evolution--- they exhibit further
exciting features as compared to photons.

One is that the violation of $CP$ symmetry in the mixing of $K^0 \bar K^0$ leads to a
violation of a Bell inequality for quasi--spin variation refuting in consequence any
local realistic theory.

Another feature is that Bell inequalities for time variations are ---due to the unitary time
evolution which includes the decay states--- much more sophisticated than in the photon case. A
CHSH--Bell inequality can be violated for a certain initial state thus ruling out local realistic
theories. This nonlocal state is not maximally entangled and shows therefore the difference of the
conceptions nonlocality and entanglement. The interesting question is how such a nonlocal state
(where the $K_S K_S$ and $K_L K_L$ parts dominate) can be produced at DA$\Phi$NE.

Furthermore, using the regeneration feature of the kaons other type of Bell inequalities can be
established. The analysis of all possible Bell inequalities together with the choice of suitable
initial states and experimental set--ups will be of great importance for testing quantum mechanics
at DA$\Phi$NE. Work in this direction is in progress\cite{DomenicoGoHiesmar}.

\section{Acknowledgements}

We would like to thank A. Di Domenico for the invitation to the Workshop on
\textit{Neutral Kaon Interferometry at a $\Phi$--Factory} at Frascati, March 2006. We
also acknowledge financial support from EURIDICE HPRN-CT-2002-00311.

\end{document}